\pdfoutput=1
\documentclass[sigconf,screen]{acmart}

\usepackage{comment}
\usepackage{diagbox}
\usepackage{algorithm,algpseudocode, algorithmicx}
\algrenewcommand\alglinenumber[1]{\tiny #1:} 
\usepackage{setspace, etoolbox, caption}
\usepackage{graphicx}
\usepackage{epstopdf}
\usepackage{listings}
\usepackage{makecell,multirow}
\usepackage{url}
\usepackage{xcolor}
\usepackage{verbatim}
\usepackage{balance}
\usepackage{pifont}
\usepackage{enumitem}
\usepackage{adjustbox}
\usepackage{amsthm}
\usepackage{amsmath}
\usepackage{amsfonts}
\usepackage{textcomp,booktabs}
\usepackage[normalem]{ulem}
\usepackage{caption,subcaption}
\usepackage{tcolorbox}
\usepackage[T1]{fontenc}
\usepackage{threeparttable}
\usepackage{multicol}
\usepackage{flushend}
\usepackage{soul}
\usepackage{xspace}
\usepackage{mathtools} 

\lstnewenvironment{tabularlstlisting}[1][]
 {%
  \lstset{aboveskip=-1.3ex,belowskip=-3.5ex,#1}%
 }
 {}

\makeatletter
\let\old@lstKV@SwitchCases\lstKV@SwitchCases
\def\lstKV@SwitchCases#1#2#3{}
\makeatother
\usepackage{lstlinebgrd}
\makeatletter
\let\lstKV@SwitchCases\old@lstKV@SwitchCases

\lst@Key{numbers}{none}{%
    \def\lst@PlaceNumber{\lst@linebgrd}%
    \lstKV@SwitchCases{#1}%
    {none:\\%
        left:\def\lst@PlaceNumber{\llap{\normalfont
                \lst@numberstyle{\thelstnumber}\kern\lst@numbersep}\lst@linebgrd}\\%
        right:\def\lst@PlaceNumber{\rlap{\normalfont
                \kern\linewidth \kern\lst@numbersep
                \lst@numberstyle{\thelstnumber}}\lst@linebgrd}%
    }{\PackageError{Listings}{Numbers #1 unknown}\@ehc}}
\makeatother

\newcommand{\ALL}{533}
\newcommand{\crash}{373}
\newcommand{\RVP}{241}

\newcommand{\incorrectoutput}{103}
\newcommand{\EO}{42}
\newcommand{\PI}{7}
\newcommand{\hang}{8}

\newcommand{\semanticserror}{156}
\newcommand{\scopeerror}{47}
\newcommand{\typecheckingerror}{61}

\newcommand{\FVerror}{75}
\newcommand{\modelcheckererror}{25}
\newcommand{\smtencodingerror}{50}
\newcommand{\codegenerror}{59}

\newcommand{\optimizationerror}{35}
\newcommand{\syntaxanalysiserror}{30}
\newcommand{\memoryissue}{44}
\newcommand{\docerror}{10}
\newcommand{\fileerror}{nine}

\newcommand{\benchmarkbugnum}{108}
\newcommand{\benchmarksolversionnum}{36}

\newcommand{\eg}{\emph{e.g.}}
\newcommand{\ie}{i.e.}

\newcommand{\compiler}{Solidity compiler\xspace}
\newcommand{\PL}{Solidity\xspace}

\newcommand\myparagraph[1]{
  \noindent \textit{\textbf{#1.}}\quad
}
\newcommand\mysubparagraph[1]{
  \noindent \textit{#1.}\,
}
\newcounter{num}
\newcommand{\takeaway}[1]{ \begin{tcolorbox}[
                  boxsep=0pt,
                  left=2pt,
                  right=2pt,
                  top=2pt,
                  bottom=2pt]
\textbf{Takeaway \refstepcounter{num}\thenum}: #1
\end{tcolorbox}}

\newif\ifletter
\letterfalse

\usepackage{cleveref} 

\Crefname{algocf}{Algorithm}{Algorithms}
\crefname{algocf}{Algorithm}{Algorithms}

\Crefname{algorithm}{Algorithm}{Algorithms}
\crefname{algorithm}{Algorithm}{Algorithms}

\crefname{appendix}{Appendix}{Appendices}
\Crefname{appendix}{Appendix}{Appendices}

\Crefname{figure}{Figure}{Figures}
\crefname{figure}{Figure}{Figures}

\crefname{listing}{Example}{Examples}
\Crefname{listing}{Example}{Examples}

\Crefname{table}{Table}{Tables}
\crefname{table}{Table}{Tables}

\crefname{thm}{Theorem}{Theorems}
\Crefname{thm}{Theorem}{Theorems}

\crefformat{chapter}{\S#2#1#3}
\crefmultiformat{chapter}{\S\S#2#1#3}{ and~#2#1#3}{, #2#1#3}{, and~#2#1#3}

\crefformat{section}{\S#2#1#3}
\crefmultiformat{section}{\S\S#2#1#3}{ and~#2#1#3}{, #2#1#3}{, and~#2#1#3}

\AtBeginDocument{%
  }

\setcopyright{acmlicensed}
\acmDOI{10.1145/3650212.3680362}
\acmYear{2024}
\copyrightyear{2024}
\acmISBN{979-8-4007-0612-7/24/09}
\acmConference[ISSTA '24]{Proceedings of the 33rd ACM SIGSOFT International Symposium on Software Testing and Analysis}{September 16--20, 2024}{Vienna, Austria}
\acmBooktitle{Proceedings of the 33rd ACM SIGSOFT International Symposium on Software Testing and Analysis (ISSTA '24), September 16--20, 2024, Vienna, Austria}
\acmSubmissionID{issta24main-p889-p}
\received{2024-04-12}
\received[accepted]{2024-07-03}

\begin{document}

\title{Towards Understanding the Bugs in Solidity Compiler}

\author{Haoyang Ma}
\orcid{0000-0002-7411-9288}
\affiliation{
    \institution{The Hong Kong University of Science and Technology\\Guangzhou HKUST Fok Ying Tung Research Institute}
    \country{China}
}
\email{haoyang.ma@connect.ust.hk}

\author{Wuqi Zhang}
\orcid{0000-0001-8039-0528}
\affiliation{
    \institution{The Hong Kong University of Science and Technology\\Guangzhou HKUST Fok Ying Tung Research Institute}
    \country{China}
}
\email{wuqi.zhang@connect.ust.hk}

\author{Qingchao Shen}
\orcid{0000-0002-6128-2123}
\affiliation{
\department{
College of Intelligence and Computing
}
    \institution{Tianjin University}
    \country{China}
}
\email{qingchao@tju.edu.cn}

\author{Yongqiang Tian}
\authornote{Corresponding authors.}
\orcid{0000-0003-1644-2965}
\affiliation{
\institution{The Hong Kong University of Science and Technology\\Guangzhou HKUST Fok Ying Tung Research Institute}
\country{China}}
\email{yqtian@ust.hk}

\author{Junjie Chen}
\orcid{0000-0003-3056-9962}
\affiliation{
\department{
College of Intelligence and Computing
}
    \institution{Tianjin University}
    \country{China}
}
\email{junjiechen@tju.edu.cn}

\author{Shing-Chi Cheung}
\orcid{0000-0002-3508-7172}
\authornotemark[1]
\affiliation{
    \institution{The Hong Kong University of Science and Technology\\Guangzhou HKUST Fok Ying Tung Research Institute}
    \country{China}
}
\email{scc@cse.ust.hk}

\begin{abstract}
\compiler{} plays a key role in enabling the development of smart contract applications on Ethereum by governing the syntax of a domain-specific language called \PL{} and performing compilation and optimization of \PL{} code.
The correctness of \compiler{}
is critical in fostering transparency, efficiency, 
and trust in industries reliant on smart contracts.
However,
like other software systems,
\compiler{} is prone to bugs,
which may produce incorrect bytecodes on blockchain platforms, 
resulting in severe security concerns.
As a domain-specific compiler for smart contracts, 
\compiler{} differs from other compilers in many perspectives, 
posing unique challenges to detect its bugs.

To understand the bugs in \compiler{}
and benefit future research,
in this paper, 
we present the first systematic study on \ALL{} \compiler{} bugs. 
We carefully examined their characteristics (including symptoms, root causes, and distribution), and their triggering test cases.
Our study leads to seven bug-revealing takeaways for \compiler{}. 
Moreover,
to study the limitations of \compiler{} fuzzers and bring our findings into practical scenarios, we evaluate three \compiler{} fuzzers on our constructed benchmark.
The results show that these fuzzers are inefficient in detecting \compiler{} bugs. 
The inefficiency arises from their failure to consider the interesting bug-inducing features, bug-related compilation flags, and test oracles. 

\end{abstract}

\ccsdesc[500]{Software and its engineering~Compilers}
\ccsdesc[500]{Software and its engineering~Software defect analysis}
\keywords{Solidity Compiler Bug, Empirical Study, Compiler Testing}
\maketitle

\section{Introduction}

Smart contracts, as self-executing agreements built on blockchain platforms like Ethereum, have revolutionized traditional contract processes by automating and enforcing the terms and conditions of agreements without the need for intermediaries. 
The significance of smart contracts lies in their ability to enhance transparency, efficiency, and trust in various domains, including finance, supply chain management, real estate, and more.
Smart contracts nowadays manage over \$47B digital assets as of November 2023~\cite{defillama}.

Existing smart contracts are mostly written in the Solidity language (over 90\%~\cite{solidityportion}) and compiled to bytecode running on an Ethereum virtual machine (EVM). 
The quality of the compiler is pivotal to the programming of smart contracts, their runtime performance and trustworthiness.
Like other compilers \cite{ChaliasosTech,EMI,zellerfragments,chengnianStudy,csmith,qingchaostudy,ChaliasosStudy,kitten,ccmd}, smart contract compiler is not immune to bugs that can affect performance, reliability, and security of the compiled contracts, even when the source code is flawlessly crafted~\cite{solcompilersecurityerror1,solcompilersecurityerror2,solcompilersecurityerror3}.
A notable reported smart contract compiler flaw was its inadvertent transformation of a well-implemented reentrancy guard at the source code level into a malfunctioning version in the bytecode. 
This concealed flaw remained undetected for over two years until it was exploited by malicious entities in July 2023, culminating in a financial loss of approximately \$73.5 million~\cite{vyperbug}.
Hence, understanding the nuances of bugs in smart contract compilers, particularly those impacting the accuracy of the compilation process and the reliability of the resulting smart contracts, is essential to safeguard the integrity of these contracts. 
Such understanding is critical in preempting and mitigating potential vulnerabilities that could compromise the security and efficacy of smart contracts in real-world applications.

Our study on the \compiler{} aims to provide a comprehensive characterization of the symptoms and root causes of its bugs. 
The \compiler{} differs from conventional compilers in several key aspects:
First, unlike traditional compilers that mainly deal with static memory allocation, the \compiler{} additionally manages on-chain persistent storage in compiled smart contracts. 
Second, a unique and primary goal of the \compiler{} is to minimize gas consumption in smart contract execution. 
Each bytecode instruction in a compiled smart contract consumes a specific amount of gas, for which users pay at runtime.
The compiler's optimization strategies are tailored to reduce this cost.
Moreover, the \compiler{} incorporates formal verification in its compilation process, a feature not commonly found in conventional compilers, to enhance the security of smart contracts.
Due to these unconventional features, the characteristics of bugs in the \compiler{} are likely distinct from those in traditional compilers. 
Conclusions drawn from existing bug studies on conventional compilers may not directly apply to the quality assurance of the \compiler{}.
Therefore, understanding the characteristics and distribution of these bugs is essential for improving the overall quality of the \compiler{} and ensuring the integrity of smart contracts.

To this end, we conduct the first complete investigation of all reported bugs in the \compiler{}. 
Specifically, we collect all 1,210 GitHub issues before the end of Oct 16 2023~\cite{solcompilerbuglist} labeled with \textit{bug}, 
exclude bugs that have no dedicated fix, and deduplicate the remaining ones. 
Finally, \ALL{} bugs are retained for further analysis. 
We highlight five \textit{symptoms} exhibited by \compiler{} bugs and delve into their twelve underlying \textit{root causes}.
We analyze the correlation between the bug symptoms and root causes. Based on the findings, we explore the challenges of triggering \compiler{} bugs and suggest potential solutions.
To gain a deeper insight into these challenges, it is also important to investigate to what extent off-the-shelf \compiler{} fuzzers address them.
Therefore, we evaluate off-the-shelf \compiler{} fuzzers in terms of bug detection. By analyzing the limitations of these fuzzers, we can identify areas where our potential bug-detection solutions can provide valuable assistance. 

To facilitate the evaluation, we build a benchmark 
consisting of \benchmarkbugnum{} reproducible bugs collected in our study and \benchmarksolversionnum{} corresponding buggy versions.
We included three publicly available 
fuzzers tailored for \compiler{} in our evaluation, including AFL-compiler-fuzzer~\cite{AFL-compiler-fuzzer}, solfuzzer~\cite{solfuzzer}, and Fuzzol~\cite{FUZZOL}. 
The evaluation results show that both of them are weak in detecting \compiler{} bugs. They can detect only a single bug in the latest compiler version and a total of 14 out of the \benchmarkbugnum{} in the benchmark.  

To sum up, we make the following major contributions.
\begin{itemize}[leftmargin=*, topsep=0pt]
    \item We conduct the first systematic study on \PL{} compiler bugs and present a classification of their symptoms and root causes. 
    \item We offer a set of seven takeaways to facilitate \compiler{} bug detection. 
    \item We develop a benchmark for the bug detection performance evaluation on \compiler{}. The benchmark is designed to be extensible and serves as a valuable resource for future evaluations of new \PL{} compiler fuzzers.
    \item We assess the bug detection performance of the existing \PL{} compiler fuzzers, 
    identify their weaknesses,
    and provide valuable insights that lay the foundation for future research to enhance the performance of bug detection for the \compiler{}.

\end{itemize}
\section{Background}
Smart contracts are self-executing agreements with terms and conditions in the form of code. They are designed to run on blockchain platforms.
In \PL{}, smart contracts are defined as classes in Object-Oriented Programming (OOP) and compiled into executable bytecodes on EVM. 
\compiler{} is responsible for carrying out this compilation process.
\begin{figure}[ht]
\vspace{-2.5mm}
\centering
    \includegraphics[width=0.75\linewidth]{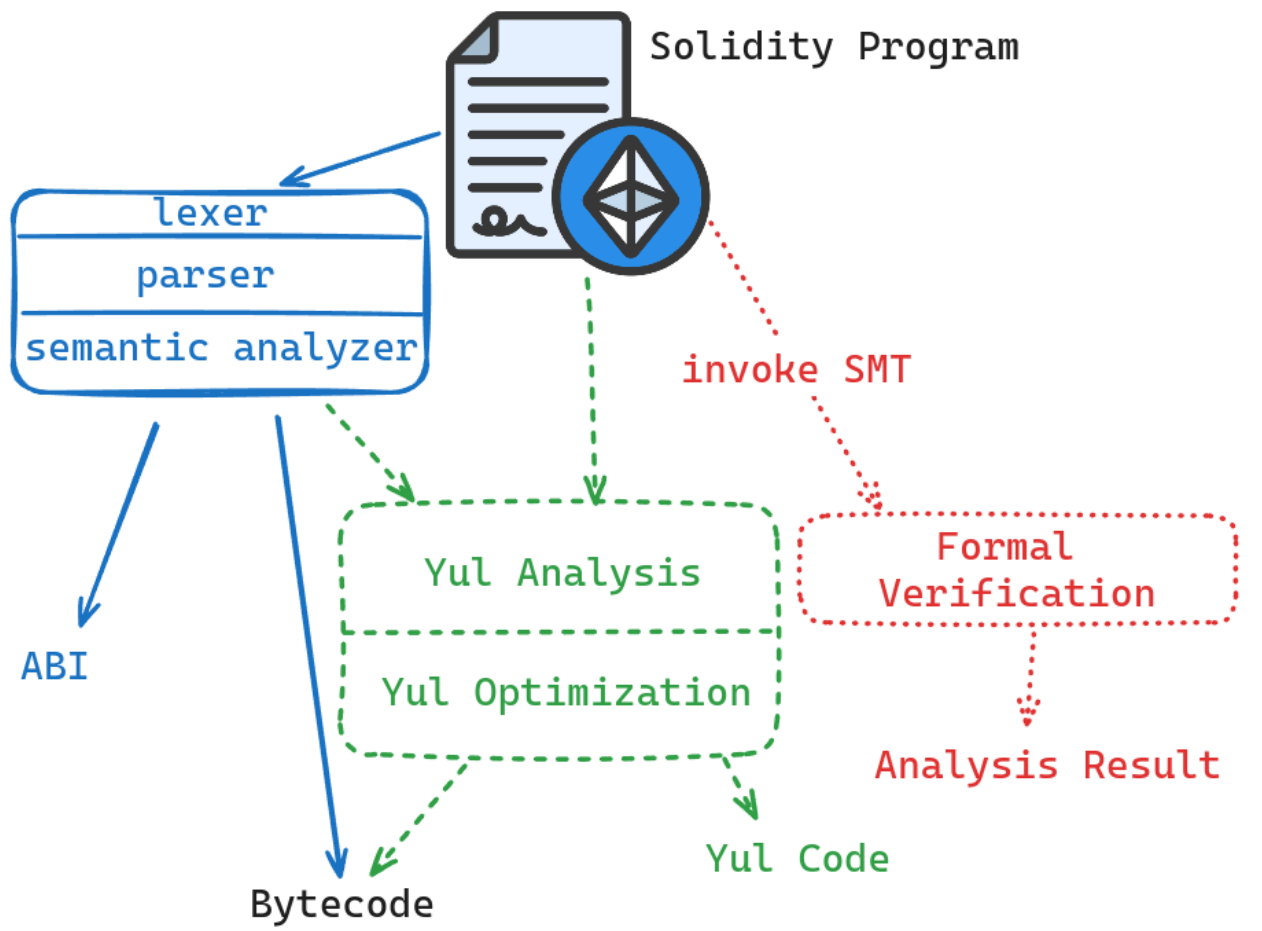}
    \caption{Workflow of \compiler{} } 
    \label{figure:compiler}
\end{figure}

\cref{figure:compiler} illustrates the workflow of \compiler{}. 
The fundamental process is represented by the blue flow in solid boxes and lines: once the \PL{} program successfully passes the frontend's checks, it undergoes conversion into bytecode that is executable on the EVM. 
Alongside the bytecode, an Application Binary Interface (ABI) is generated
for the contract. 
This ABI enables external entities to comprehend how to interact with the functions
and events of this contract.
The green flow in dashed boxes and lines involves Yul, an intermediate language that can be compiled to bytecode for different backends. Yul comes into play when the \PL{} program includes inline assembly or when specific compilation flags (\eg, \texttt{--optimize-yul}) are used to explicitly invoke Yul.
The red flow in dotted boxes and lines focuses on formal verification within \compiler{}. As a security-oriented tool, the compiler offers model checkers that assist in conducting automated mathematical proofs, ensuring that the \PL{} program adheres to a specific formal specification. This verification process enhances the overall security and reliability of the program.
The Yul and formal verification components can be optionally activated using compilation flags.
All these flows are prone to bugs. 
We will introduce them in \cref{section:bs}.
\section{Bug Collection and Classification}

\begin{figure}[ht]
\centering
    \includegraphics[width=0.9\linewidth]{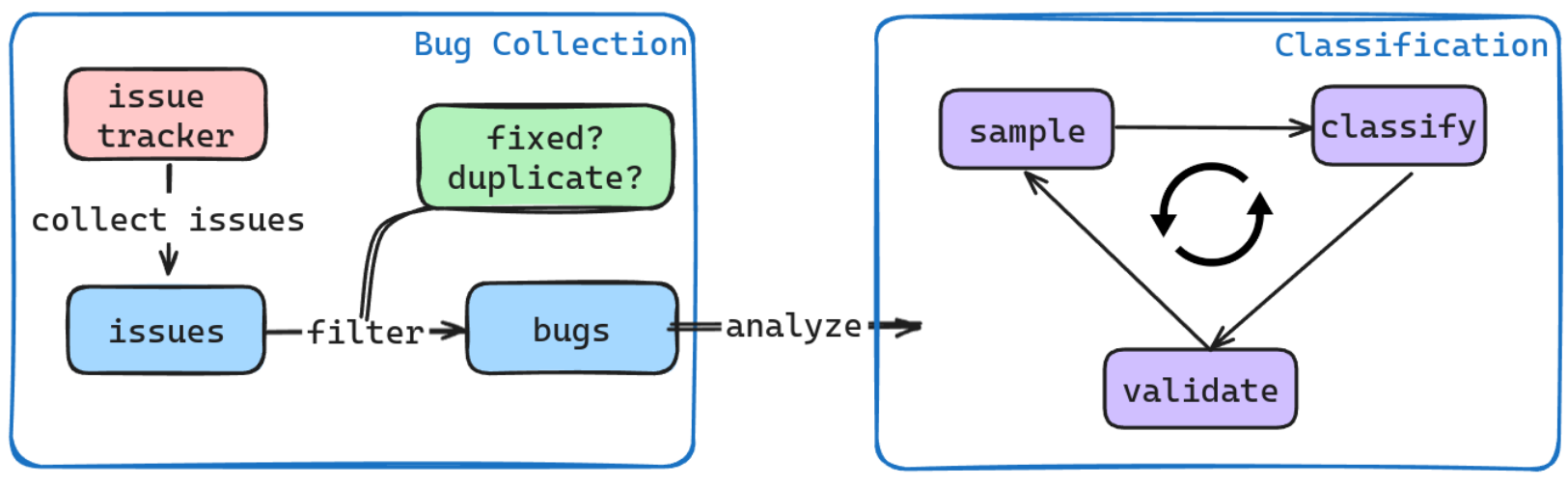}
    \caption{The Overview of Bug Collection and Classification} 
    \label{figure:overview}
\end{figure}

To conduct this study,
we first collect the bugs from \compiler{}
and classify them based on their symptoms and root causes.
This section introduces this procedure
and details the results.
\cref{figure:overview} shows the overview of this procedure.

\subsection{Bug Collection}
We collect \compiler{} bugs by extracting all closed issues tagged as ``bug'' by \compiler{} developers.
In total, we obtain a collection of 1,210 issues on GitHub. 
Then, we exclude unresolved and misreported issues from the dataset as they do not reveal any actual bugs.
We also exclude issues from our dataset that are duplicate reports for the same bug in the compiler.
Furthermore, to effectively understand the root cause of bugs from their code patches, we remove the GitHub issues where bugs are inadvertently fixed by unrelated commits.
Additionally, we exclude issues involving the collective patching of multiple bugs, as it is difficult to isolate each bug and its corresponding patch for root cause analysis in our study.
In total, \ALL{} unique
 bugs are obtained for subsequent analysis. 

\subsection{Bug Classification}
We classify the collected bugs in multiple iterations following the open coding practice~\cite{qualitativeanalysis}. 
In each iteration, 50 bugs are randomly sampled and two researchers independently examine them and label the symptoms and root causes of each bug.
Afterward, the two researchers jointly validate their results by comparing and discussing the differences in their labels. 
To increase accuracy, the validation process includes a third researcher with expertise in Solidity smart contracts. 
After finishing the validation, a new iteration begins on 50 bugs that have not been analyzed yet.
This iteration continues until a consensus is reached on all the labels for the \ALL{} bugs.
Following existing works~\cite{qingchaostudy, dlsysbugstudy1}, we observe Cohen’s Kappa coefficient~\cite{5584447} to measure the inter-rater agreement. After labeling the first 50 bugs, the Cohen’s Kappa coefficient is close to 0.
After discussing the disagreement and the validation process, the coefficient rises to 72\% after labeling the next 50 bugs and remains constantly above 95\% after labeling 150 bugs.


\begin{figure}[ht]
\centering
    \includegraphics[width=0.7\linewidth]{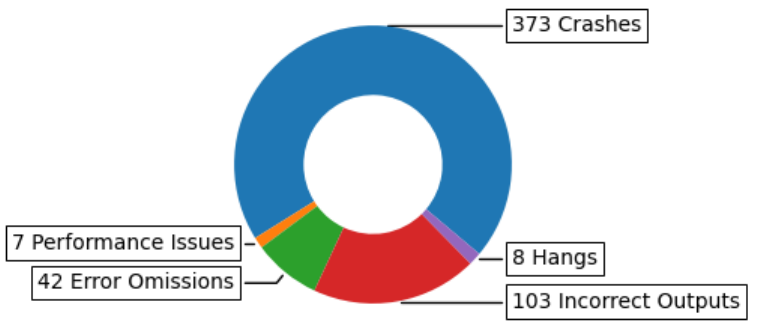}
    \caption{Bug Distribution Over Symptoms}
    \label{figure:symptoms}
\end{figure}

\subsubsection{Symptoms}
\label{section: symptoms}
In total, there are five symptoms in all the collected \PL{} compiler bugs. 
The bug distribution over symptoms is presented in \cref{figure:symptoms}. 

\myparagraph{Symptom 1: Crash}
\label{section:crash}
Crash refers to an unexpected termination or abnormal termination of a program during its execution. It manifests as the sudden halting of the program's functionality, often accompanied by error messages or system notifications. Crashes can occur due to various reasons, such as memory leaks, resource exhaustion, etc. When a crash occurs, it signifies a failure in the software's stability and reliability. Like other compilers\cite{qingchaostudy,ChaliasosStudy,csmith}, the crash symptom is the major issue among all collected bugs in \compiler{}, occupying \crash{} out of \ALL{} bugs. More granularly speaking, the crash symptom consists of two sub-symptoms: 1) \textit{Valid Program Rejection}  and 2) \textit{Uninformative Error Message}. 

\definecolor{stubbg}{HTML}{FCF3D5} 
\begin{figure}[h!]
\lstinputlisting[
language=c,
morekeywords={TypeError, unit, immutable, contract, function, int, public},
label={listing:RVP},
caption={Valid Program Rejection (Issue \#12379~\cite{RVP})},
basicstyle=\scriptsize
]{
CODE/RVP.tex
} 
\end{figure}

\mysubparagraph{Valid Program Rejection}
Given a valid test program following \PL{}'s grammar, 
\compiler{} rejects it with an informative error message (\eg, \cref{listing:RVP}). 
This symptom takes up \RVP{} out of \crash{} crash bugs. 

\definecolor{stubbg}{HTML}{FCF3D5}
\begin{figure}[h!]
\lstinputlisting[
language=c,
label={listing:PEM},
caption={Uninformative Error Message (Issue \#13101~\cite{PEM})},
basicstyle=\scriptsize
]{
CODE/PEM.tex
}
\end{figure}
\vspace{-8mm}
\definecolor{stubbg}{HTML}{FCF3D5}
\begin{figure}[h!]
\lstinputlisting[
language=c,
label={listing:PEM2},
caption={Informative Error Message After Fixing},
basicstyle=\scriptsize
]{
CODE/PEM2.tex
}
\end{figure}

\mysubparagraph{Uninformative Error Message}
An uninformative Error Message refers to the crash symptom where the thrown error message is not informative enough for debugging.
In other words, the crash should occur due to errors in the contract code, but users of the compiler desire a more informative error message to help them troubleshoot.
As shown in \cref{listing:PEM}, \compiler{} only reports that the crash is triggered by a \texttt{bad\_cast} without extra information for debugging, e.g.,~where the bad cast occurs in the contract code. 
After fixing this compiler bug, \compiler{} can pinpoint the crash-triggered line in the contract as well (\cref{listing:PEM2}).

\definecolor{stubbg}{HTML}{FCF3D5}
\begin{figure}[h!]
\lstinputlisting[
language=c,
label={listing:incorrect_output},
morekeywords={contract, function, uint, public, private, library, using, for},
caption={Incorrect Output (Issue \#10170~\cite{incorrect_output5})},
linebackgroundcolor = {\ifnum \value{lstnumber} > 2 \ifnum \value{lstnumber} < 4 \color{stubbg} \fi \fi}, 
numbers=left,
basicstyle=\scriptsize
]{
CODE/incorrect_output.tex
}
\end{figure}

\myparagraph{Symptom 2: Incorrect Output}
Incorrect Output refers to the symptom where the output generated by \compiler{} deviates from the expectation. The outputs include SMT-LIB Script~\cite{incorrect_output1}, helper/warning messages~\cite{incorrect_output2}, bytecodes~\cite{incorrect_output3}, returned values~\cite{incorrect_output4}, Application Binary Interface (ABI)~\cite{incorrect_output5}, etc. 

\cref{listing:incorrect_output} shows an example of incorrect ABI generation.
The following code is a \PL{} contract that defines two functions: \texttt{emptyStringRevert} and \texttt{test}. The \texttt{emptyStringRevert} function is a pure function that reverts the transaction with an empty string as the revert reason. The \texttt{test} function is a view function that calls the \texttt{emptyStringRevert} function using a try-catch statement and returns the revert reason and its length as a tuple of bytes and uint256.
ABI is a standard way to interact with contracts in the Ethereum ecosystem, both from outside the blockchain and for contract-to-contract interaction. The revert reason in the above code is ABI-encoded as if it were a call to a function Error (string).
\compiler{} performs the incorrect ABI encoding and generates an unexpected extra 32 bytes, which corrupts the interaction.


\myparagraph{Symptom 3: Error Omission}
Error omission refers to a situation where an expected error or exception condition is not properly detected, reported, or handled within a software system. In the context of \compiler{}, it incorrectly remains silent and does not report any error when receiving an ill-formed test program. There are \EO{} bugs of this symptom.

\myparagraph{Symptom 4: Performance Issue}
A performance issue refers to a situation where the bytecode generated by \compiler{} does not meet the expected or desired level of performance in terms of speed, responsiveness, or resource utilization due to improper optimizations performed by \compiler{}.
There are \PI{} performance issues, and 5 of them are due to optimization errors. 

\myparagraph{Symptom 5: Hang}
A hang symptom refers to a situation where a software system or application becomes unresponsive or frozen. \hang{} bugs exhibit this symptom.

\begin{figure}[ht]
\centering
    \includegraphics[width=0.9\linewidth]{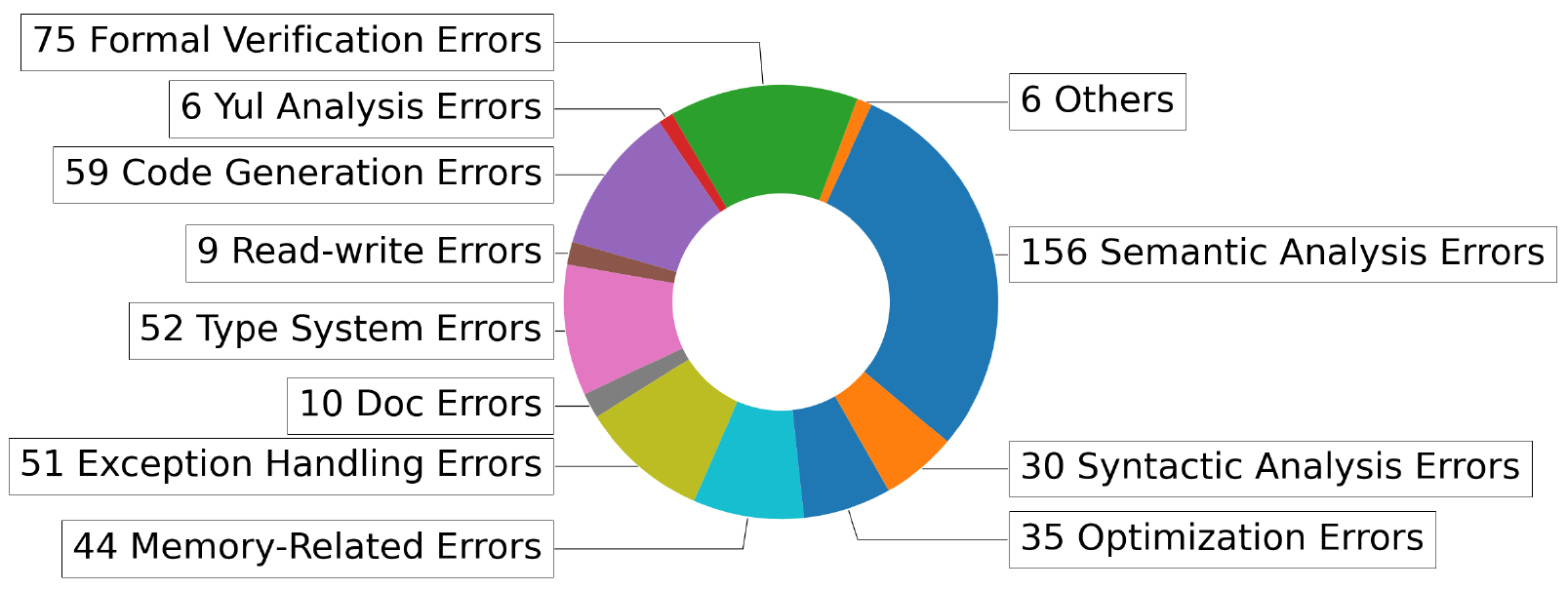}
    \caption{Bug Distribution Over Root Causes.
    } 
    \label{figure:rootcauses}
\end{figure}

\subsubsection{Root Causes}
\label{section: rootcauses}

In total, all the bugs can be grouped into 12 classes of errors. 
The name of each error implies the root cause.
The frequency of each root cause is presented in \cref{figure:rootcauses}. 
Formal verification errors and Yul analysis errors are unique to \compiler{}.
Though other root causes spread around other compilers, \compiler{} brings its flavors into them.
For instance, the special memory management system on blockchain platforms induces special memory-related keywords and features in \compiler{}, triggering \memoryissue{} bugs.
Such flavors introduce new challenges in bug-revealing that cannot be addressed by previous studies.
We will discuss them in \cref{section: RQ1}.
Below is a detailed description of all the root causes, in descending order of their frequency of occurrence.

\myparagraph{Root Cause 1: Semantic Analysis Error}
Semantic analysis goes beyond checking the syntax and focuses on extracting knowledge about the program from built-up contexts (\eg, type system, control flow), understanding its intended behavior, and verifying it.
For instance, given a type-contained abstract syntax tree (AST) node, the semantic analyzer consults the type system implementation about its expected type(s) and compares the actual type with the expected one(s). This process, conventionally named type checking~\cite{typecheckingerror}, is a part of semantic analysis. 
In addition, the analyzer also cares for variables' scopes~\cite{scopeerror}, object-oriented programming (OOP) features~\cite{OOPerror}, control flow~\cite{controlflowerror}, etc.

\definecolor{stubbg}{HTML}{FCF3D5} 
\begin{figure}[h!]
\lstinputlisting[
language=c,
morekeywords={contract, function, uint, public, private, library, using, for},
label={listing:semanticanalysiserror},
caption={Semantic Analysis Error (Issue \#13764~\cite{scopeerror})},
escapechar=|,
linebackgroundcolor = {\ifnum \value{lstnumber} > 3 \ifnum \value{lstnumber} < 5 \color{stubbg} \fi \fi}, 
numbers=left,
basicstyle=\scriptsize
]{
CODE/semanticanalysiserror.tex
} 
\vspace{-5mm}
\end{figure}

\mysubparagraph{Example Bug}
\cref{listing:semanticanalysiserror} reveals a silent semantic analysis error. In particular, on line 4, variables of type \textit{uint} are mistakenly granted access to the function \texttt{privateFunction}, thereby exposing the private function's scope beyond the intended private field. However, the compiler throws no error message on this misbehavior.

\myparagraph{Root Cause 2: Formal Verification Error}
Security issues inside the compiled smart contracts can result in costly consequences, such as irreversible economic losses~\cite{EthBMC,ZEUS,10.1145/3460319.3464837}.
Therefore, identifying security issues in the \PL{} code is significant. To this end, \compiler{} provides a formal verification component to perform model checking on the to-be-compiled code. 
The error that happened inside this component while being related to the model checking process is named formal verification error.

\definecolor{stubbg}{HTML}{FCF3D5} 
\begin{figure}[h!]
\lstinputlisting[
language=c,
morekeywords={contract, function, uint, public, private, library, using, for},
label={listing:fverror},
caption={Formal Verification Error (Issue \#14111~\cite{fverror})},
escapechar=|,
linebackgroundcolor = {\ifnum \value{lstnumber} > 11 \ifnum \value{lstnumber} < 14 \color{stubbg} \fi \fi
\ifnum \value{lstnumber} > 5 \ifnum \value{lstnumber} < 8 \color{stubbg} \fi \fi
\ifnum \value{lstnumber} > 1 \ifnum \value{lstnumber} < 4 \color{stubbg} \fi \fi
}, 
numbers=left,
basicstyle=\scriptsize
]{
CODE/fverror.tex
} 
\end{figure}

\mysubparagraph{Example Bug}
\ifletter
Example 6 
\else
\cref{listing:fverror} 
\fi
detects a formal verification error in a constrained horn clauses (CHC) model checker.
Specifically, 
\textit{guard} is initialized with true (line 3), and \textit{v} is assigned an initial value 0 implicitly (line 2).
The value of \textit{guard} is switched to false on line 12, and the call to \texttt{this.dec} will decrease the value of the unsigned int \textit{v} by one (line 7) since \textit{guard} is false (line 6). However, CHC fails to build a call edge from \texttt{f} to \texttt{dec} in call graph building, and thus does not report the underflow error.

\myparagraph{Root Cause 3: Code Generation Error}
Code generation is the process of automatically generating code based on a set of rules and specifications. Within \compiler{}, the code generation process involves producing several formats, including IR, EVMASM (Ethereum Virtual Machine Assembly), bytecode, WebAssembly (Wasm), and JSON-format AST.

\definecolor{stubbg}{HTML}{FCF3D5} 
\begin{figure}[h!]
\lstinputlisting[
language=c,
morekeywords={contract, function, uint, public, private, library, using, for},
label={listing:codegenerror},
caption={Code Generation Error (Issue \#3271~\cite{codegenerror})},
escapechar=|,
linebackgroundcolor = {\ifnum \value{lstnumber} > 4 \ifnum \value{lstnumber} < 8 \color{stubbg} \fi \fi
}, 
numbers=left,
basicstyle=\scriptsize
]{
CODE/codegenerror.tex
} 
\end{figure}

\mysubparagraph{Example Bug}
\cref{listing:codegenerror} 
detects a code generation error when generating bytecode. Specifically, the bytecode for the internal function \texttt{f} is incorrectly included in the deployed bytecode, which abuses the visibility of \texttt{f}.



\myparagraph{Root Cause 4: Type System Error}
Like other general programming languages and domain-specific languages (DSLs)~\cite{ChaliasosStudy}, \PL{}'s type system is ideally sound in theory but practically bug-prone in implementation~\cite{soliditypatchstudy}. 
Specifically, the type system defines allowable types in \PL{}, permissive operations on these types, and type relations (\eg, whether type conversion or type coercion is allowed between two types). Beyond these hardcoded rules, in addition to its primary functions, a type system within a compiler has the capability to perform type inference, allowing it to deduce the resulting type after an operation is performed.

\definecolor{stubbg}{HTML}{FCF3D5} 
\begin{figure}[h!]
\lstinputlisting[
language=c,
morekeywords={contract, function, int, public},
label={listing:typesystemerror},
caption={Type System Error (Issue \#9548~\cite{typeerror2})},
escapechar=|,
linebackgroundcolor = {\ifnum \value{lstnumber} > 3 \ifnum \value{lstnumber} < 5 \color{stubbg} \fi \fi}, 
numbers=left,
basicstyle=\scriptsize
]{
CODE/typesystemerror.tex
} 
\end{figure}

\mysubparagraph{Example Bug}
\cref{listing:typesystemerror} shows a bug-triggered test case for a typical type system error. The yellow strip above the code snippet highlights the line where the bug was triggered. In this instance, \compiler{} forgets to disallow the use of the negative number type during type inference for the resultant type of exponentiation.


\myparagraph{Root Cause 5: Exception Handling Error}
The exception-handling component is responsible for responding to unexpected events during execution. This component is helpful for bug localization and guidance. Therefore, errors in exception handling can confuse the users and complicate the bug isolation process for developers.

\definecolor{stubbg}{HTML}{FCF3D5} 
\begin{figure}[h!]
\lstinputlisting[
language=c,
morekeywords={contract, function, uint, public, private, library, using, for},
label={listing:exceptionhandlingerror},
caption={Exception Handling Error (Issue \#11610~\cite{exceptionhandlingerror})},
escapechar=|,
linebackgroundcolor = {\ifnum \value{lstnumber} > 4 \ifnum \value{lstnumber} < 6 \color{stubbg} \fi \fi
}, 
numbers=left,
basicstyle=\scriptsize
]{
CODE/exceptionhandlingerror.tex
} 
\end{figure}

\mysubparagraph{Example Bug}
\cref{listing:exceptionhandlingerror} reveals an exception handling error. Specifically, line 5 includes an internal type \textit{S} in a public function, which disrespects the visibility of \textit{S} and should trigger an informative error reporting, but it does not. The error is a misleading message like "std::exception::what: [solidity::util::tag\_comment*] =". After a code patch, \compiler{} can correctly report "TypeError 4103: (56-69): Internal type is not allowed for public or external functions" to complain about the incorrect use of \texttt{S[2]}.



\myparagraph{Root Cause 6: Memory-Related Error}
A memory-related error in a compiler refers to problems related to allocating, managing, and using memory during the compilation process. These issues can arise from various factors, such as excessive memory consumption, leaks, or inefficient memory utilization. Memory-related errors can lead to compiler crashes, slow performance, or unexpected outputs during compilation.

\begin{figure}[h!]
\lstinputlisting[
language=c,
morekeywords={contract, function, uint, public, private, memory, using, for},
label={listing:memoryissue},
caption={Memory-Related Error (Issue \#12558~\cite{memoryissue})},
escapechar=|,
linebackgroundcolor = {\ifnum \value{lstnumber} > 7 \ifnum \value{lstnumber} < 9 \color{stubbg} \fi \fi
}, 
numbers=left,
basicstyle=\scriptsize
]{
CODE/memoryissue.tex
} 
\end{figure}

\mysubparagraph{Example Bug}
\cref{listing:memoryissue} detects a memory copy error.
\texttt{sStorage} is a variable that reside in \textit{storage}, a long-term storage space available in the Ethereum blockchain.
\textit{sMemory} is a variable located in \textit{memory}, a temporary storage area used during the execution of functions.
Line 8 performs copying storage structs to memory arrays, leading to a memory copy error that is manifested as a crash.

\myparagraph{Root Cause 7: Optimization Error}
Optimization on smart contracts can efficiently reduce both the size of the deployed bytecode and execution cost, \ie, the required gas for contract deployment and calls to the contract. As a special optimization target, gas is an artificially designed measurement of computation to prevent denial-of-service attacks and ensure that the Ethereum network remains secure and efficient.
The existence of gas makes the optimization in \compiler{} unique and requires special care.
Incorrect optimization may lead to compilation crashes or even incorrect bytecode generation.

\definecolor{stubbg}{HTML}{FCF3D5} 
\begin{figure}[h!]
\lstinputlisting[
language=c,
morekeywords={contract, function, uint, mstore, private, library, using, for},
label={listing:optimizationerror},
caption={Optimization Error (Issue \#6246~\cite{optimizationerror})},
escapechar=|,
linebackgroundcolor = {\ifnum \value{lstnumber} > 4 \ifnum \value{lstnumber} < 6 \color{stubbg} \fi \fi
}, 
numbers=left,
basicstyle=\scriptsize
]{
CODE/OptimizationError.tex
} 
\end{figure}

\mysubparagraph{Example Bug}
\cref{listing:optimizationerror} is a Yul code that reveals an optimization error.
The optimizer should have simplified line 5 to \texttt{mstore(0,0)}, but instead, it reduces it to \texttt{mstore(0,2)}. 
\texttt{mstore} is the instruction that stores a value (the $2_{nd}$ argument) to a memory location (the $1_{st}$ argument). 
Since the inferred value of a bitwise shift expression (\texttt{shl(1,shl(not(0),x))}) is incorrect, the optimized instruction stores the wrong value to memory location $0$ and leads to generating incorrect bytecode.

\myparagraph{Root Cause 8: Syntactic Analysis Error}
Syntax refers to a sequence of characters or tokens that are intended to be written in a particular programming language. Compilers leverage grammar to constrain the occurrences of tokens. Incorrect syntax analysis may lead to accepting programs that violate the grammar or early compilation crashes.
In total, we collect \syntaxanalysiserror{} syntactic analysis errors.

\definecolor{stubbg}{HTML}{FCF3D5} 
\begin{figure}[h!]
\lstinputlisting[
language=c,
morekeywords={contract, interface, event, public, indexed, library, using, for},
label={listing:syntaxanalysiserror},
caption={Syntactic Analysis Error (Issue \#13681~\cite{syntaxerror})},
escapechar=|,
linebackgroundcolor = {\ifnum \value{lstnumber} > 2 \ifnum \value{lstnumber} < 4 \color{stubbg} \fi \fi
}, 
numbers=left,
basicstyle=\scriptsize
]{
CODE/syntaxanalysiserror.tex
} 
\vspace{-5mm}
\end{figure}

\mysubparagraph{Example Bug}
\cref{listing:syntaxanalysiserror} shows a syntactic analysis error. Multiple indexed entries are syntactically invalid, but the compiler omits this error.

\myparagraph{Root Cause 9: Doc Error}
Natspec is short for \textit{Natural Specification}, which refers to a documentation system used in \PL{} smart contracts.
Natspec allows developers to write human-readable documentation and explanatory comments directly within the \PL{} code.
Errors contained in Natspec are named \textit{doc error}, and the number of them is \docerror{}.

\myparagraph{Root Cause 10: Read-write Error}
\compiler{} can communicate with operating systems by reading from/writing into source files and input/output buffers. In total, \fileerror{} bugs are about read-write errors.

\myparagraph{Root Cause 11: Yul Analysis Error}
Similar to the code analysis for the \PL{} source codes, the Yul component also provides analysis for Yul code.
Six bugs are contained in the Yul component when analyzing the Yul code.

\myparagraph{Root Cause 12: Others}
The root causes of six bugs are rare and do not fall into any other root cause. 
\section{Bug Analysis}
\label{section:bs}

This section aims to answer the following research questions:

\noindent\textbf{RQ1 (Bug Characteristics) What is the correlation between symptoms and root causes? How do all bugs distribute over the three compiler components?} 
The correlation between symptoms and root causes reveals how compiler bugs manifest themselves.
Understanding it is crucial for effective problem diagnosis, troubleshooting, and efficient resolution of issues.  
Besides, the bug-proneness of different compiler components may be different. 
Investigating the bug distribution over components offers insights for testing on one specific component.

\noindent\textbf{RQ2 (Test Case Characteristics) What are the primary characteristics of bug-revealing test cases?}
Bug-revealing test cases for a specific root cause may contain common characteristics, which may guide effective compiler testing.

\noindent\textbf{RQ3 (Test Oracle Requirement) What test oracles are required for revealing bugs?}
Test oracles help reveal bugs triggered by test cases.
A non-negligible portion (30.02\%) of our collected \ALL{} bugs does not result in a crash. 
Designing effective test oracles is essential to identify many compiler bugs.
\subsection{RQ1: Bug Characteristics}
\label{section: RQ1}
\begin{table*}[ht]
\centering
\caption{Correlation between Symptoms and Root Causes}
\begin{adjustbox}{width=0.95\textwidth,center}
\begin{tabular}{l|c|c|c|c|c|c|c}
\toprule
\multirow{2}{*}{\diagbox{Root Causes}{Symptoms}} & \multicolumn{2}{c|}{Crash}                             & \multirow{2}{*}{Incorrect Output} & \multirow{2}{*}{Error Omission} & \multirow{2}{*}{Performance Issue} & \multirow{2}{*}{Hang} & \multirow{2}{*}{$\textit{Total}_{\textit{symptom}}$} \\
\cline{2-3}
                   & Valid Program Rejection & Uninformative Error Message &  & & & & \\
  \midrule
Semantic Analysis Error                    & 56 & 61 & 14 & \boxed{23}  & 1 & 1 & 156 \\
Formal Verification Error           & 59 & 4 & 7 & 5  & - & - & 75 \\
Code Generation Error            & 27 & 6 & \boxed{24} & 2  & - & - & 59  \\
Type System Error                      & 27 & 13 & 7 & 4  & - & 1 & 52  \\
Exception Handling Error    & 18 & \boxed{24}  & 8 & 1  & - & - & 51  \\
Memory-Related Error                & 22  & 12  & 9 & 1 & - & - & 44  \\
Optimization Error                 & \boxed{8} & -  & 15 & 2  & 5 & \boxed{5} & 35  \\
Syntactic Analysis Error            & 12 & 5  & 9 & 3  & 1 & - & 30  \\
Doc Error & 5 & 2 & 3 & -  & - & - & 10  \\
Read-write Error                          & 3 & 3  & 3 & -  & - & - & 9  \\
Yul Analysis Error                            & 3 & 1  & 1 & 1  & - & - & 6  \\
Others                     & 1  & 1  & 3 & -  & - & 1 & 6  \\ \midrule
\textit{$\textit{Total}_{\textit{causes}}$}                         & 241 & 132 & 103 & 42 &7 & 8 & 533 \\ \bottomrule
\end{tabular}
\end{adjustbox}
\label{tab:symptom_causes}
\end{table*}

\subsubsection{Correlation between Symptoms and Root Causes}
\label{section: correlation}

\cref{tab:symptom_causes} presents the correlation between symptoms and root causes. We highlight several interesting figures with wrapped boxes around the numbers. From these figures, we have the following findings.

\myparagraph{Error Omission and Semantic Analysis Error}
Among the \EO{} bugs categorized under the error omission symptom, a significant majority, accounting for 54.76\%, are attributed to semantic analysis errors. The remaining error omissions are evenly distributed across other root causes.
This observation implies that when localizing uninformative and hard-to-pinpoint error omissions, it can be beneficial to initially narrow down the search to files associated with semantic analysis.


\myparagraph{Incorrect Output and Code Generation Error}
Different from other root causes, triggering a code generation error requires a careful inspection of the generated code. Among \codegenerror{} collected bugs, 24 are of symptom Incorrect Output. Code generation errors have a significantly higher proportion of Incorrect Output symptom compared to the average ($38.98\%$ \textit{v.s.} $18.91\%$), implying that the detection of bugs under this root cause requires special test oracles on the generated code.

\myparagraph{Crash and Exception Handling Error}
82.35\% of exception-handling errors manifest themselves as crashes, while 57.14\% of such crashes are actually \textit{Uninformative Error Message}, implying that more than half of the crashes caused by exception-handling errors are expected to happen, but should be accompanied by more informative messages.

\myparagraph{Crash and Optimization Error}
Due to the fact that improper optimizations tend to affect the generated bytecodes instead of raising crashes,
only 22.86\% of optimization errors display crash symptoms, much lower than the average proportion (\ie, 69.98\%) of this primary symptom.
Therefore, special test oracles are required for the generated output, performance, and running time.

\myparagraph{Hang and Optimization Error}
While hang is not typically considered a core symptom of bugs in \compiler{}, it can be troublesome in practice due to its lack of informative details. Hangs often occur without providing a stack trace or error message, making it challenging to isolate and address them effectively. However, it is worth noting that among the \hang{} hangs observed, 5 of them are caused by optimization errors. This observation can potentially aid in localizing and pinpointing the causes of hangs, offering valuable insights for bug localization and resolution in such scenarios.

\begin{figure}[ht]
\centering
    \includegraphics[width=0.90\linewidth]{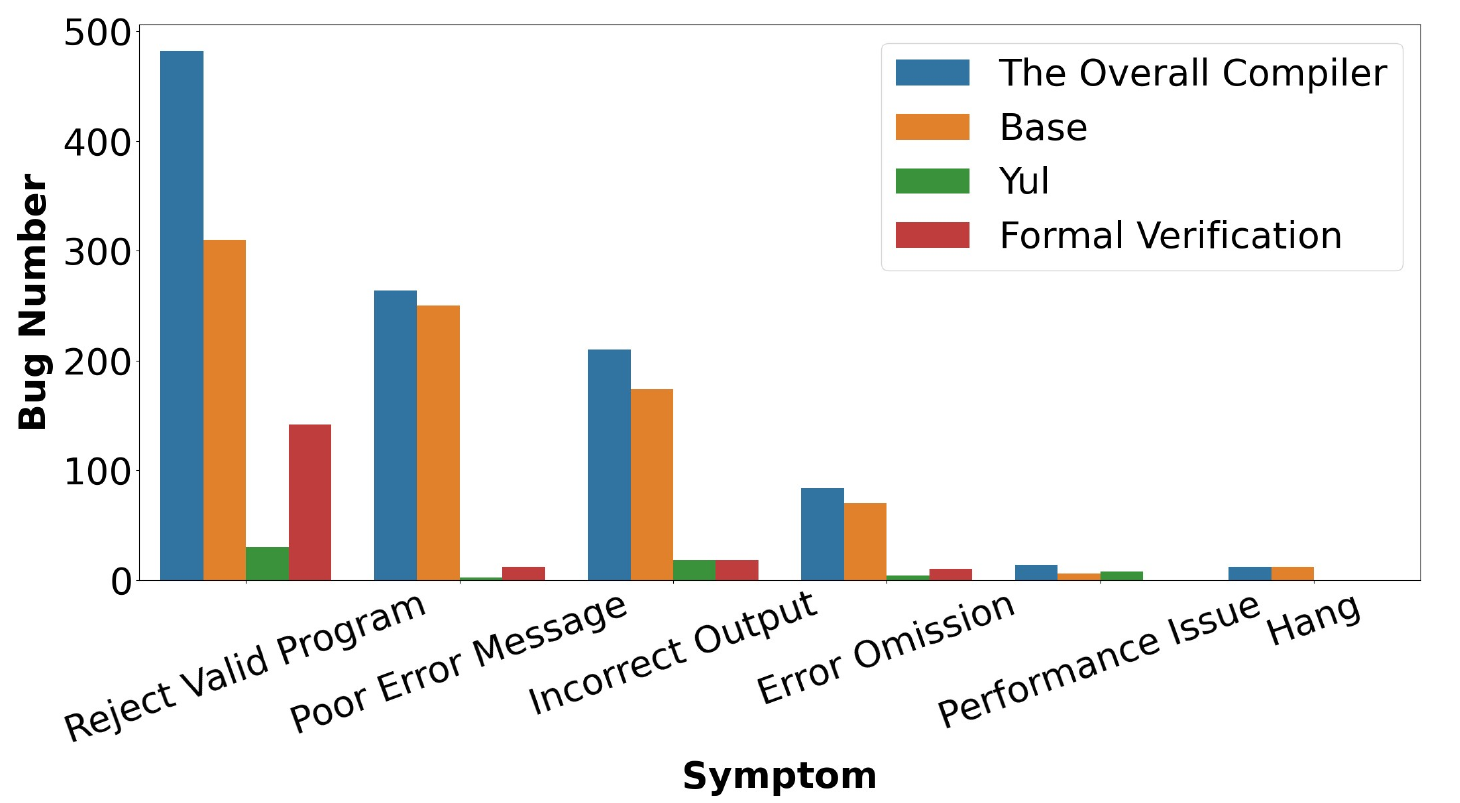}
    \caption{Bug Distribution Over Components}
    \label{figure:components}
\end{figure}

\subsubsection{Bug Distribution over Components}
\cref{figure:components} shows the bug distribution over symptoms in three different components shown in \cref{figure:compiler}.
Several conclusions can be drawn from this figure.
\begin{itemize}[leftmargin=*, topsep=0pt]
    \item The primary symptom observed across all components is the crash symptom (\ie, Valid Program Rejection plus Uninformative Error Message). However, the proportion of this symptom varies among the different components.
    \item The distribution of bugs across symptoms in the base component is comparable to that observed in the entire compiler.
    \item In the formal verification component, the symptom of rejecting valid programs is considerably more prevalent compared to others, occupying 78.02\% of all the bugs inside this component. This observation suggests that bug detection in formal verification may not require highly specialized test oracles.
    Nevertheless, to effectively conduct fuzzing on this component, it is crucial to take into account additional bug-exposing factors (\cref{section: additional-factor}).
\end{itemize}

In addition to symptoms, bugs in different components have distinct underlying causes. 
All root causes are present in the base component,
while in the Yul component, the major root causes are Yul analysis error and optimization error, accounting for 70.97\% of all bugs. 
In the formal verification component, the primary root cause is formal verification error, constituting 82.42\% of all bugs. Semantic analysis error also occurs in this component due to the requirement of semantic analysis for encoding by the SMT encoder.

\takeaway{Formal Verification, being a distinctive component of the compiler, demonstrates distinct patterns in the distribution of bug symptoms and root causes.}

\subsection{RQ2: Test Case Characteristics}
\label{sec: RQ2}
\subsubsection{Bug-Revealing Code Patterns}
This section presents four bug-revealing code patterns that are usable for detecting specific errors.

\myparagraph{Trigger Semantic Analysis Error}
Of the \semanticserror{} collected semantic analysis errors, it is worth mentioning that \scopeerror{} of them pertain to scope issues, including the bug triggered by \cref{listing:semanticanalysiserror}. In their bug-revealing tests, several scope-defining and scope-changing features are used, such as 1) the use of \textit{private}, \textit{public}, and \textit{external} in function declaration; 2) the involvement of OOP-related features; 3) the use of \textit{using for} directive to attach in new functionalities to a specific type; 4) function overloading; and 5) variable shadowing.
Another prominent subcategory is the type-checking issue. The semantic analyzer verifies the correctness of node information in the AST by checking statements. Type checking, such as a simple \texttt{variable.isLibraryFunctionParameter()}, should be actively incorporated to handle corner cases, unsuggested behavior, and misbehavior. However, type-checking issues, such as the absence of a required checking, lead to \typecheckingerror{} bugs out of \semanticserror{} bugs under semantic analysis errors. The combinatory use of types (\eg, mapping from a public view function to a struct) is probable to detect such bugs.

\takeaway{Valid stacking of Solidity's scope-related features (\eg, qualifiers such as \textit{external}, scope changer such as \textit{using for}, etc) and type combinations is beneficial for detecting semantic analysis errors.}

\myparagraph{Trigger Formal Verification Error}
The incorrect model checker implementation, as shown in \cref{listing:fverror}, occupies \modelcheckererror{} out of \FVerror{} bugs under Formal Verification Error root cause. The bug-triggered test inputs for such bugs contain well-designed and obscure data flows that require complete analysis. Besides the implementation problem in model checkers, the rest \smtencodingerror{} bugs are located in the SMT encoder. The SMT encoder encodes solidity expressions into SMT expressions for model checkers to resolve. Specifically, the encoder first decomposes a solidity expression into pieces and then generates a suitable SMT abstraction for each piece. Finally, the decoder integrates these SMT abstractions into complete SMT expressions. The encoder errors lie in the decomposition and generation phases, triggered by difficult-to-encode solidity expressions, such as arrays, array slices, tuples, and mappings. These four elements are present in 33 out of \smtencodingerror{} bugs.
\takeaway{Well-designed data flows plus the presence of key elements in Solidity (static array, dynamic array, array slice, tuple, and mapping) is important for detecting Formal Verification Error.}

\myparagraph{Trigger Memory-Related Error}
Memory-related keywords, features, and operations are the key to triggering memory-related errors.
In \PL{}, keywords such as \texttt{memory}, \texttt{calldata}, features such as \texttt{dynamic array}, \texttt{inline assembly} about direct memory management (\eg, \texttt{mstore}), operations such as defining large-size arrays and out-of-bound member access can all induce memory-related errors. 36 out of \memoryissue{} memory-related errors are triggered by the above elements. The remaining eight bugs are about null pointer dereferences during implementation.

\takeaway{
Memory-related keywords (\eg, \textit{memory, calldata}) and operations (\eg, memory copy between memory locations defined by memory-related keywords) are required for triggering memory-related errors.}

\myparagraph{Trigger Optimization Error}
Among \optimizationerror{} optimization errors, nearly half ($16$) leads to incorrect bytecode generation. Similar to the code generation error, special test oracles are required for detecting optimization errors by inspecting the generated bytecode. 
Besides, triggering optimization errors requires the use of Yul code as input. 18 out of \optimizationerror{} optimization errors are revealed by Yul code.
\takeaway{
Yul code is required for efficiently detecting optimization errors.
We can involve it in Solidity's inline assembly or in an independent input file.
}

\subsubsection{Bug-Revealing Compilation Flags}
\label{section: additional-factor}
Triggering special bugs (\eg, bugs of incorrect output symptom, bugs inside the formal verification component, and the Yul component) require special cautions, such as traversing compilation flags.
\compiler{} provides compilation flags to invoke specialized code generations or to involve specific compiler components.
For instance, 
\texttt{--asm} and \texttt{--asm-json} are flags for generating EVM assemblies;
\texttt{--ir}, \texttt{--ir-ast-json},\\ \texttt{--ir-optimized}, and  \texttt{--ir-optimized-ast-json} are flags for generating Yul codes;
\texttt{--abi} is used to produce ABIs.
When designing a fuzzer to detect bugs characterized by incorrect output symptoms, these flags become necessary.
To invoke the Yul component, \texttt{--optimize-yul},  \texttt{--yul-optimizations} are required.
Regarding formal verification, the compiler provides 13 compilation flags to customize the verification processes. 
Intelligently utilizing these flags and traversing different combinations with them is important 
for specific bug detection in \compiler{}.

\takeaway{Special compilation flags are required for testing formal verification component and Yul component.}

\subsection{RQ3: Test Oracle Requirement}
\label{sec: RQ3}
This section discusses the test oracles required to detect special \compiler{} bugs.

\myparagraph{Output-Aware Test Oracle}
 Among \ALL{} bugs, \incorrectoutput{} bugs are of Incorrect Output symptom. Such a symptom occurs quietly without an unexpected termination and thus needs special care.
 For instance,
as is implied by \cref{section: correlation}, an output-aware test oracle is required for effectively detecting code generation errors and optimization errors since such bugs are highly related to incorrect compilation output.

\myparagraph{Error-Message-Aware Test Oracle}
 About 24.76\% of \ALL{} bugs are of Uninformative Error Message symptom. This particular symptom distinguishes itself from a typical crash by being an expected occurrence. The fixes for such bugs are about refining error messages by providing more bug information or changing the severity level. 
 For instance, as is suggested by \cref{section: correlation}, an error-message-aware test oracle is required for revealing exception handling errors efficiently.
 
\takeaway{To effectively resolve 24.76\% of \ALL{} bugs, an error-message-aware test oracle is required to determine whether the error message is misleading, or whether the error message lacks debugging-required information (\eg, stack trace).}

\myparagraph{Validity-Aware Test Oracle}
Since there are \EO{} bugs of Error Omission symptom, a validity-aware test oracle is needed for such bugs by judging the validity of the \PL{} code and catching silent response from \compiler{} to an invalid test input.

\section{Evaluation on \compiler{} Fuzzers}
\label{sec: fuzzer}

Following the discoveries in the bug analysis 
\ifletter
(Section 4),
\else
(\cref{section:bs}), 
\fi 
it is vital to examine the effectiveness of off-the-shelf \compiler{} fuzzers and delve into their limitations. This investigation plays a crucial role in comprehending how we can leverage our insights to enhance the efficacy of fuzzing strategies.
To this end, this section presents the evaluation process and analyzes potential improvements for Solidity compiler fuzzers.



\subsection{Fuzzer Selection}

The primary criterion for selecting a fuzzer revolves around whether it is specifically designed for \compiler{} or tailored to suit its specific features.
We have chosen three modified general fuzzers, namely AFL-compiler-fuzzer~\cite{AFL-compiler-fuzzer}, solfuzzer~\cite{solfuzzer}, and Fuzzol~\cite{FUZZOL}. 
These fuzzers are variations of AFL~\cite{AFL}, customized and adapted to enhance fuzzing efficiency for \compiler{}. 
To uphold the principle of fairness in this experiment, we unify the hyperparameter MAP\_SIZE in the three fuzzers. MAP\_SIZE limits the number of Instrumentations and thus influences coverage results. We configure MAP\_SIZE as 16, aligning with the suggestion from AFL.

\myparagraph{AFL-compiler-fuzzer}
AFL-compiler-fuzzer aids AFL with several language-agnostic mutation rules (\eg, rewriting conditions in \texttt{if} statements, removing statements, etc) specifically for conventional code structures.
Besides, AFL-compiler-fuzzer can compose code pieces into new test cases to both enrich seed pool and reuse bug-revealing code fragments. 

\myparagraph{solfuzzer}
solfuzzer is an open-source fuzzer integrated within \compiler{}.
It uses the AFL executable to carry out the fuzzing process.
During the fuzzing loop, solfuzzer extends the capabilities of AFL by trying out different compilation flags, 
in order to cover new code paths and increase the bug-revealing possibility.
It is proved that several bugs only exhibit with the normal solc executable and can not be detected by the mere AFL~\cite{solfuzzer-exp}.

\ifletter

\else
\myparagraph{Fuzzol}
\fi
Unlike AFL-compiler-fuzzer and solfuzzer, Fuzzol stands out by incorporating language-specific mutators. 
It begins by analyzing the ASTs of Solidity test cases and then applies the mutators to AST nodes. For example, Fuzzol can substitute an AST node with another node of the same type or modify the arguments of an opcode within the inline assembly node.
This approach enables Fuzzol to provide more targeted and language-aware mutations to enhance the effectiveness of the fuzzing process.
\subsection{Benchmark Construction}
To evaluate the performance of the three \PL{} compiler fuzzers, we construct a benchmark using all the bugs collected by this work. 
Each bug in the benchmark is associated with four elements: bug description, code patch, test cases, and the version of the compiler that introduced the bug.
To facilitate compiler fuzzer evaluation, for each bug, we use the nearest buggy release version as the subject for fuzzing test.
We collect both the test cases and the compiler version if the bug is reproducible on the buggy version. 
However, we only consider reproducible bugs with crash symptoms in our fuzzer evaluation benchmark
since the three fuzzers are not equipped with test oracles to identify other symptoms, such as 
analyzing the generated bytecode or ABI, determining the expected error messages, analyzing poor performance, or detecting runtime hangs. 
In total, we collect \benchmarkbugnum{} reproducible bugs and \benchmarksolversionnum{} buggy release versions. 

\subsection{Seed Selection}
Fuzzers have their own preferences for seeds. 
For instance, Fuzzol prefers small test cases with less than 1KB in size, while AFL-compiler-fuzzer prefers code templates (\ie, code snippets with placeholders) and code fragments (\ie, the concrete representation of placeholders in Solidity)
extracted from test cases. 
In the experiment, we evaluate these fuzzers
in two settings.
In the first setting, 
for each fuzzer, we used the seed collection procedure 
mentioned in its documentation or original paper. 
Specifically, 
for AFL-compiler-fuzzer, we collect test cases from \compiler{} and exclude test cases that trigger bugs in the benchmark.
Then, we employ \textit{comby-decomposer}~\cite{combydecomposer} to decompose these test cases into fine-grained templates and fragments for invoking the code composition of AFL-compiler-fuzzer.
We obtain 5,812 test cases, 25,305 fragments, and 39,978 templates.
On average, each test case contains 32.9 lines of code (LOC).
The seed pool construction for solfuzzer can be accomplished by executing \textit{isolate\_tests}, which is an automatic test case collector for solfuzzer. 
In total, 308 test cases are collected for solfuzzer, each containing 41.0 LOC.
As for Fuzzol, we follow its experimental setup
and collect seeds from its public repository~\cite{fuzzolexp}. 
In total, 1,466 test cases are collected for Fuzzol, 
each containing only 13.6 LOC on average.
In the second setting, we use the same seeds for all the three fuzzers. 
Specifically, 
we choose 
5,812 test cases for AFL-compiler-fuzzer to build the common seed pool for the three fuzzers.
This seed pool includes all unit test cases for the latest version of \compiler{} except for bug-revealing test cases in the benchmark. 
We consider the number of bug-revealing test cases generated from this shared seed pool as a metric to assess the bug-detection abilities of each fuzzer concerning bugs in the benchmark.

\subsection{Fuzzing Loop}
Our fuzzing loop is built around the latest version of \compiler{}.
In each fuzzing iteration, each fuzzer selects a seed and mutates it in order to generate new test input. This process aims to enhance code coverage on the latest version of \compiler{}.
In particular, we turn on the comby server for AFL-compiler-fuzzer to splice fragments and templates in the seed pool.
As the fuzzing target, the latest version plays two roles.

\myparagraph{A bug-contained individual}
As a bug-contained individual,
the number of crashes inside the latest version identified by these three fuzzers serves as a robust indicator of their effectiveness in uncovering bugs in \compiler{}.

\myparagraph{A fuzzing proxy}
Given the fact that the collected \benchmarkbugnum{} bugs are distributed across \benchmarksolversionnum{} compilers of distinct versions, instrumenting all of these compilers and carrying out the fuzzing loops on each of them independently is resource-consuming and time-consuming.
To address this practical challenge, we adopt a strategy known as "fuzzing by proxy" inspired by recent advancements in fuzzing techniques~\cite{Donaldsonissta23, fuzzbyproxy1, fuzzbyproxy2}. 
The proxy refers to a typical software under test that can guide the mutation in the fuzzing loop to generate a representative set of test inputs for all similar software.
In our approach, the proxy is the latest version of \compiler{}.
The efforts paid by the three fuzzers on increasing code coverage on the latest version of \compiler{} help expand the diversity of the generated mutants.
After collecting a swarm of mutants,
we examine the bug-revealing ability of the generated mutants on the collected buggy versions.

\begin{table}[t]
    \centering
    \small
    \caption{Fuzzing Result}
    \vspace{-2mm}
    \label{tab:fuzzing}
    \begin{tabular}{c|c|c|c}
         \toprule
         \textbf{Fuzzer} & \textbf{\#Crash}$_l$  & \textbf{\#Bug}$_b$ & \textbf{Map Density} 
         \\
         \midrule
         AFL-compiler-fuzzer & 1 & 8 & 80.12\% ($\Uparrow6.11\%$) \\
         solfuzzer & 0 & 0 & 88.22\% ($\Uparrow3.70\%$)\\
         Fuzzol & 0 & 5 & 81.00\% ($\Uparrow13.02\%$)\\
        Fuzzol$_{acf}$ & 0 & 1 & 80.45\% ($\Uparrow6.44\%$)\\
        solfuzzer$_{acf}$ & - & - & -\\
        \bottomrule
    \end{tabular}
        \begin{tablenotes}
    \footnotesize
    \item \textbf{\#Crash}$_l$: the number of revealed crashes in the latest version of \compiler{}
    \item \textbf{\#Bugs}$_b$: the number of detected bugs inside the benchmark
    \item \textbf{Map Density}: Map density indicates how many branch tuples we have hit. It is in proportion to edge coverage~\cite{AFL++}
    \item \textbf{Fuzzol}$_{acf}$ and \textbf{solfuzzer}$_{acf}$: Fuzzol and solfuzzer that fuzzes from AFL-compiler-fuzzer's seeds
    \end{tablenotes}
\end{table}
\subsection{Fuzzing Result}
\cref{tab:fuzzing} 
presents the fuzzing results of 20-day executions on a server with AMD Ryzen Threadripper 3970X 32-Core Processor, 
and 256G RAM, coordinated
with 64-bit Ubuntu 22.04 OS.

In the first experimental setting, where each fuzzer utilizes its preferred seeds, AFL-compiler-fuzzer ranks first by uncovering eight bugs.
Fuzzol follows closely in second place, detecting five bugs. solfuzzer does not find any bug in this particular setting. 
While utilizing the seeds of AFL-compiler-fuzzer, Fuzzol detects a bug that is previously undetected by the AFL-compiler-fuzzer. 
However, Fuzzol does not detect this particular bug using its preferred seeds. 
On the other hand, solfuzzer encounters difficulties in loading these seeds and ultimately reports a timeout.
Note that bugs detected by Fuzzol under two experimental settings do not overlap with bugs detected by AFL-compiler-fuzzer, suggesting that these two effective fuzzers explore distinct opportunities for revealing bugs.
We further explore the root causes of bugs detected by the AFL-compiler-fuzzer and Fuzzol under two experimental settings.
Positively, these fuzzers perform well in detecting memory-related errors.
Out of 12 such errors collected in the benchmark, five have been revealed. 
However, they are comparably inefficient in detecting semantic analysis errors and formal verification errors, which are the two major root causes that constitute 43.40\% of the overall 533 bugs studied.
Out of 24 semantic analysis errors and 36 formal verification errors collected in the benchmark, the three fuzzers could detect only one semantic analysis error and five formal verification errors. 
Detection of semantic analysis errors requires precise and meticulous mutators to manipulate fine-grained program elements like scope, as highlighted in takeaway 2. However, neither Fuzzol nor AFL-compiler-fuzzer incorporates mutators of this nature, such as lifting the scope of a variable. To detect formal verification errors, fuzzers should focus on modifying data flows, especially among contracts, since the formal verification component aims at inspecting data and revealing data corruption issues, such as integer overflow. Both fuzzers lack specific mutators to achieve this modification. 

\ifletter
\textbf{Enhancing Bug Detection}

\else
\subsection{Enhancing Bug Detection}
\fi
Findings and insights in our study can benefit fuzzers from the following aspects.
First, as indicated by takeaways 
\ifletter
2, 3, and 4 in Section 4.2, 
\else
2, 3, and 4 in \cref{sec: RQ2}, 
\fi
the proposition of designing mutators based on scope, type, and data flow, along with generating varied code snippets incorporating these elements, is beneficial in detecting semantic analysis errors and formal verification errors. 
These two categories are identified as the primary root causes of bugs in \compiler{}. 
A program synthesizer generating such code snippets can enrich error-revealing seeds and thus invoke fuzzers' bug-detection effectiveness.

Furthermore, the three chosen baselines all consider crash as the test oracle. However, code generation errors, exception handling errors, and optimization errors often manifest as non-crashes or uninformative-error-messages 
\ifletter
(as explained in Section 4.1.1).
\else
(according to \cref{section: correlation}).
\fi
Hence, fuzzers should examine error messages, warning messages, and outputs to effectively detect \compiler{} bugs. 
Moreover, \compiler{} fuzzers can learn from partial compiler fuzzers~\cite{hirgen,opera} that specialize in examining special components or compilation phrases.
Our findings on bug distributions among compiler components aid fuzzers in effectively allocating their power budget.

\section{Discussion}


\ifletter
\textbf{6.1 \compiler{} Bugs \textit{v.s.} Bugs in Other Compilers}
\else
\subsection{Bugs in \compiler{} \textit{v.s.} Bugs in Other Compilers}
\fi

\label{section: comparison}
In this section, we compare \compiler{} bugs with the bugs in other popular compilers, including C/C++ compilers~\cite{chengnianStudy, optimizationstudy,chengnianStudy}, and deep learning compilers ~\cite{qingchaostudy}. 
We make three major observations to inspire follow-up research in Solidity compiler testing.

\ifletter
\else
\begin{figure}[ht]
\centering
    \includegraphics[width=0.90\linewidth]{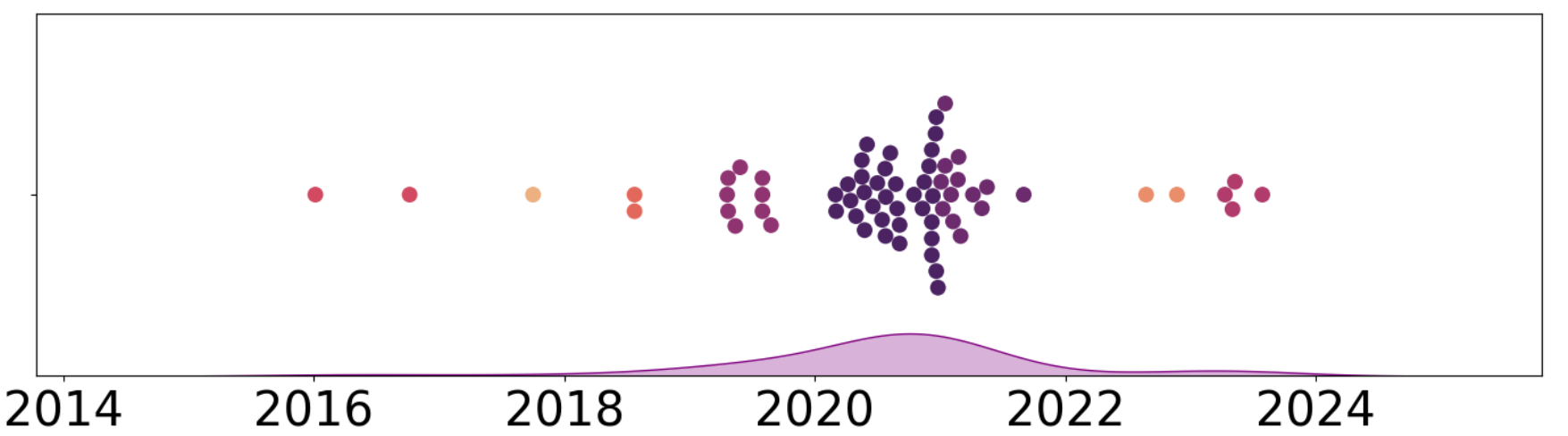}
    \caption{The Distribution of Formal Verification Errors over Years and Its Kernel Density Estimate} 
    \label{figure:fv}
\end{figure}
\fi

\myparagraph{Formal Verification Component}
The formal verification component is unique to the Solidity compiler and does not exist in other compilers like C or deep learning compilers.
Such unique component comes with unique bugs that may not occur in other compilers. 
In our study, we found that formal verification error accounts for a significant portion of all the bugs, indicating its prevalence and importance. 
Noteworthy, we also found that the number of formal verification error reports has increased rapidly since 2020, corresponding to the emerging blockchain hacks recently 
\ifletter
(Figure~\ref{figure:fv-response}).
\else
(Figure~\ref{figure:fv}).
\fi
The formal verification errors contribute to 31.57\% of the bugs collected in the last year (Oct/16/2022 - Oct/16/2023). 
The percentage is more than twice the ratio (14.07\%) of formal verification errors among all bugs. 
This indicates that security issues in smart contracts have raised more attention to the formal verification feature of the \compiler{}. 
However, the formal verification component still has much room for improvement.
Due to the absence of a formal verification component, bug knowledge and testing knowledge for the existing compilers does not apply to \compiler{}. 
Our study offers insights and guidance for the testing of the formal verification component in the compiler (Section~4.2~and~4.3). 
We call for more future research on improving the quality of the formal verification component in \compiler{}.

\myparagraph{Inline Assembly}
Inline assembly is widely used in smart contracts and \compiler{} often suffers from relevant bugs.
According to our findings, 11.63\% \PL{} bugs are triggered by inline assembly. 
Bugs induced by the inline assembly are not well studied in the literature of traditional compilers (\eg, GCC/LLVM).
Our study fills this gap by categorizing the types of compiler bugs that are more likely to be triggered with inline assembly (\eg, the memory-related error and optimization error), and offers insights into more effective test input generation of \compiler{} (\cref{sec: RQ2}).

\myparagraph{Gas Optimization}
Optimizations in \compiler{} significantly differ from other compilers. 
It is designed to minimize gas consumption, a key metric for evaluating the computational effort required for smart contract execution. 
Disparities in the motivations behind optimizations between \compiler{} and other compilers can potentially invalidate the application of prior optimization bug findings~\cite{optimizationstudy} to the optimization bugs in \compiler{}.
Our study uncovers the connection between optimization errors and symptoms like crashes or hangs, which have not been extensively discussed in the current literature.

\myparagraph{Summary}
The disparities between \compiler{} and other compilers hinder transferring the knowledge of conventional compiler bugs to \compiler{}.
Our study brings a new perspective beyond the existing ones on bugs in traditional compilers and deep learning compilers.

\ifletter
\textbf{Solidity Compiler Bugs and Security Issues}

\else
\subsection{Solidity Compiler Bugs and Security Issues}
\fi
The main objective of \compiler{} is to compile smart contracts written in Solidity into bytecodes executable on Ethereum Virtual Machine (EVM).
In the Ethereum blockchain, every node launches EVM to execute bytecodes that are derived from smart contract compilation.
Unsafe bytecode emitted by \compiler{}
can cause security issues in the blockchain.
Take a hashing error~\cite{securitybug2} on the list of known security-relevant bugs~\cite{securitybug} as an example. 
The legacy optimizer in \compiler{} can save runtime costs by computing an instruction in the inline assembly region at compile time and caching the result. The cached value will be reused directly if the same instruction occurs subsequently. This optimization reduces redundant computation but also invokes a hashing error when computing two Keccak-256 hashes of two data with the same effective content, while one of them has extra zero bytes. 
The second hash value is unexpectedly the same as the first one though their hashing contents are of different lengths. 
Attackers can use this bug to bypass security checks that are forged from these two hash values. 
Therefore, the solidity community has assigned a medium severity tag to this bug.
It is noteworthy that
all the 77 security-relevant bugs in this list are non-crash bugs. 
Among them, 24 require the use of legacy optimizer, Yul optimizer, and ABI encoder/decoder for reproduction.
Our findings in non-crash symptoms and calls for test oracles for detecting them
\ifletter
(Section 4.1, 4.3)
\else
(\cref{section: RQ1}, \cref{sec: RQ3})
\fi
encourages new research efforts to address potential security issues induced by \compiler{}.

\ifletter
\textbf{Guidance for beyond Bug Detection}

\else
\subsection{Guidance for beyond Bug Detection}
\fi

Besides guidance for fuzzing \compiler{} and uncovering its bugs, this study also provides hints for employing compiler validation approaches and program analysis on \compiler{}.
In compiler verification, our findings can integrate with Equivalence Modulo Inputs (EMI)~\cite{EMI}.
For example, Takeaway 4 emphasizes the significance of memory-related keywords and operations in detecting memory-related errors. 
This insight helps establish memory-related ground truth, verifying that data transfer between storage and memory on the blockchain does not impact data values. 
We can verify this ground truth on the \compiler{} by comparing the values of the same data in two compiled smart contracts that only differ in memory-related operations. 
For analyzing compilers, our findings about bug distribution over components help mitigate the scalability issues of program analyzers on heavyweight compilers, letting them allocate more power to bug-rich components.

\subsection{Threats to Validity}
A threat to external validity lies in the representativeness of the collected bugs.
The representativeness issue may be caused in two ways: 1) inadequate bugs or poor sampling strategy and 2) too many unanalyzable bugs in the collection.
Therefore,
to minimize the external threat, 
we consider all available GitHub issues labeled with ``bug'' in \compiler{} repository.
In addition,
We only collect bugs associated with a detailed description or discussion and a dedicated bug fix.
In this way, all the collected bugs are conducive to manual analysis and further classification.
%
The threat to internal validity mainly lies in the manual inspection, analysis, and categorization.
The categorization is subjective to the analyzer and thus can introduce bias and even errors~\cite{qingchaostudy}.
To mitigate and even eliminate this threat, we conduct an iterative cross-validation strategy between two experienced researchers and also introduce another Blockchain expert for conducting extra discussions on disagreements and executing finalization.
The nomenclature for symptoms and root causes generally follows previous compiler bug studies ~\cite{qingchaostudy, ChaliasosStudy} and employs a concise and consistent naming scheme to accurately represent bug characteristics that are exclusive to \compiler{} (\eg, Formal Verification Error).

\section{Related Work}

Understanding bug characteristics is crucial for various purposes, including automated testing~\cite{concurrencybugstudy,dlsysbugstudy2,integeroverflowstudy,chen2023toward,tensorflowbugstudy,chengnianStudy,androidbugstudy} and debugging~\cite{zeller2002simplifying,zhang2023ppr,zhang2024lpr,xu2023pushing,wang2021probabilistic,sun2018perses,chen2020enhanced,rcc}. 
Empirical bug studies primarily focus on investigating and extracting these bug characteristics to enhance the understanding of the bugs under consideration.

Certain literature concentrates on comprehending specific types of bugs.
For instance, Lu et al.~\cite{concurrencybugstudy} characterize concurrency bugs.
Franco et al.~\cite{numericalbugstudy} and Wang et al.~\cite{dlsysbugstudy2} study real-world numerical bugs.
Dietz et al.~\cite{integeroverflowstudy} investigate integer overflow bugs.
Moreover,
certain literature is dedicated to examining bugs within a particular real-world software system.
Jia et al.~\cite{tensorflowsysbugstudy} analyzed bug patterns inside the Tensorflow System.
Chen et al.~\cite{chen2023toward} carried out a comprehensive bug study for four popular DL libraries.
Garcia et al.~\cite{vehiclebugstudy} conducted a bug study on autonomous vehicles.
Islam et al.~\cite{dlsysbugstudy1}, Wang et al.~\cite{dlsysbugstudy2}, and Zhang et al.~\cite{tensorflowbugstudy} studied bugs in deep learning systems.
Wan et al.~\cite{blockchainsysbugstudy} investigated bugs in blockchain systems.

The most relevant works to our studies are the compiler bug studies.
Certain compiler bug studies investigate the \textit{whole} compiler.
For instance, Sun et al.~\cite{chengnianStudy} conducted a comprehensive study on LLVM and GCC, analyzing their bugs in several aspects including duration, priority, code patches, locations, etc. 
Shen et al.~\cite{qingchaostudy} and Du et al.~\cite{maleistudy} analyzed deep learning compiler bugs and expose their important features such as symptoms, root causes, locations, etc.
Other compiler bug studies focus on bugs in a \textit{compilation stage}.
For instance, 
Chaliasos et al.~\cite{ChaliasosStudy} concentrated on analyzing type-related compiler bugs in four JVM-based compilers.
Zhou et al.~\cite{optimizationstudy} studied the optimization-related bugs inside LLVM and GCC.
This paper focuses on bugs inside the whole \compiler{} and it is different from previous works.
As shown in \cref{figure:compiler}, \compiler{} have different structures than traditional compilers (\eg, LLVM and GCC) and deep learning compilers. 
These distinctions introduce a novel bug distribution and unique bug characteristics (\cref{section: comparison}). This paper reveals these distinctions and further provides practical guidelines for future bug detection and isolation within the compiler. 

\section{Conclusion}
This paper introduces the first empirical study on \ALL{} compiler bugs, examines their properties, presents seven key findings for efficient bug detection, and assesses three \compiler{} fuzzers. The study specifically focuses on labeling bug symptoms, summarizing their underlying causes, and exploring their interconnections. 
Additionally, the research delves into test case characteristics and the requirements for effective test oracles, providing practical recommendations for testing. The evaluation of the fuzzers involves a 20-day experiment for each fuzzer, analyzing their limitations and suggesting practical improvements based on the findings.
\section{Data Availability}
We release the bug classification results, benchmarks, experimental data, and the instructions for replicating the evaluation on \url{https://github.com/haoyang9804/ISSTA24-Solidity-Study}.

\begin{acks}
This work is supported by the National Natural Science Foundation of China (Grant No. 61932021), Hong Kong Research Grant
Council/General Research Fund (Grant No. 16205722), National Natural Science Foundation of China Grant Nos. 62322208, 62232001, and CCF Young Elite Scientists Sponsorship Program (by CAST).
\end{acks}
\balance
\bibliographystyle{ACM-Reference-Format}
\bibliography{reference}

\end{document}